\documentclass[preprint,amsmath,amssymb,aps,showkeys,showpacs]{revtex4}
\usepackage[english]{babel}
\usepackage{graphicx}
\usepackage{graphics}
\usepackage{amsmath}
\usepackage{dcolumn}
\usepackage{amssymb}
\usepackage{bm}
\usepackage[latin1]{inputenc}

\begin{document}
\title{Rotating effects on the Landau quantization for an atom with a magnetic quadrupole moment}
\author{I. C. Fonseca}
%\email{itallo.fisica@gmail.com}
\affiliation{Departamento de F\'isica, Universidade Federal da Para\'iba, Caixa Postal 5008, 58051-970, Jo\~ao Pessoa, PB, Brazil.}

\author{K. Bakke}
\email{kbakke@fisica.ufpb.br}
\affiliation{Departamento de F\'isica, Universidade Federal da Para\'iba, Caixa Postal 5008, 58051-970, Jo\~ao Pessoa, PB, Brazil.}

\begin{abstract}
Based on the single particle approximation [V. F. Dmitriev {\it et al}, Phys. Rev. C {\bf50}, 2358 (1994); C.-C. Chen, Phys. Rev. A {\bf51}, 2611 (1995)], the Landau quantization associated with an atom with a magnetic quadrupole moment is introduced, and then, rotating effects on this analogue of the Landau quantization is investigated. It is shown that rotating effects can modify the cyclotron frequency and breaks the degeneracy of the analogue of the Landau levels.  
\end{abstract}


\maketitle

\section{Introduction}

The Landau quantization \cite{landau} arises from the interaction between an electric charged particle and a uniform magnetic field perpendicular to the plane of motion of the particle. As a result of this interaction, the motion of this quantum particle acquires distinct orbits and the energy spectrum of this system becomes discrete and infinitely degenerate. At present days, the Landau quantization has attracted interests in studies of Bose-Einstein condensation \cite{l5,l6}, two-dimensional surfaces \cite{l2,l3,l4} and the quantum Hall effect \cite{l1}. Another point of view has arisen from the challenge to building a quantum system where the quantum Hall effect for neutral particles could be observed. This model has been proposed in Ref. \cite{er}, where the electric field that interacts with the permanent magnetic dipole moment of the neutral particle must satisfy specific conditions: the absence of torque on the magnetic dipole moment of the neutral particle, the electric field must satisfy the electrostatic conditions, and there exists the presence of a uniform effective magnetic field given by $\vec{B}_{\mathrm{eff}}=\vec{\nabla}\times\vec{A}_{\mathrm{eff}}$, where $\vec{A}_{\mathrm{eff}}=\vec{\sigma}\times\vec{E}$ corresponds to an effective vector potential, $\vec{E}$ is the electric field and $\vec{\sigma}$ are the Pauli matrices. Therefore, by choosing an electric field that satisfies the above conditions, it has been shown in Ref. \cite{er} that the motion of the neutral particle acquires discrete orbits, where the energy levels correspond to the analogue of the Landau levels. This analogue of the Landau levels corresponds to the Landau-Aharonov-Casher quantization \cite{er}. Other analogues of the Landau quantization have been discussed in Refs. \cite{lin,lin2,lin3,basu,fb2,fb3}.

On the other hand, noninertial effects was explored from a geometrical point of view by Landau and Lifshitz \cite{landau2}, where it is shown that the coordinate system becomes singular at large distances in a uniformly rotating frame, which is associated with the velocity of the particle would be greater than the velocity of light. From a quantum mechanical point of view, this spatial restriction pointed out by Landau and Lifshitz \cite{landau2} has been discussed in an analogue of the Landau quantization for neutral particles \cite{b}. Besides, studies of quantum mechanical systems in rotating frames have shown effects associated with the geometric quantum phases, such as the Sagnac effect \cite{sag,sag5}, the Mashhoon effect \cite{r4} and the Aharonov-Carmi geometric phase \cite{ac2}. It is worth mentioning studies made in the context of condensed matter, where rotating effects have been in the rotating Bose-Einstein condensation in ultra cold diluted atomic gases \cite{cond2}, in the quantum Hall effect \cite{cond1}, the Aharonov-Carmi geometric phase in C60 molecules \cite{cond3}, in quantum rings \cite{r12,r11,dantas} and spintronics \cite{spint1,spint2,spint3}.

The aim of this work is to analyse quantum effects on an analogue of the Landau quantization associated with an atom with a magnetic quadrupole moment due to rotating effects.  A great deal of work can be found in the literature with respect to studies of quadrupole moments of atoms and molecules in the context of the chemical physics, for instance, in single crystals \cite{quad}, refractive index \cite{quad2}, nuclear quadrupole interactions \cite{nucquad,nucquad2,quad4,quad14,quad16}, molecules \cite{quad-1,quad-2,quad3,quad5,quad7,quad8,quad9,quad10,quad11,quad12,quad13,quad15,quad17}, atoms \cite{quad6,prlquad} and superposition of chiral states \cite{quad18}. Moreover, quadrupole moments have been investigated in the context of quantum physics, such as in studies of geometric quantum phases \cite{chen,b7} and noncommutative quantum mechanics \cite{nonc}. In particular, quantum particles with a magnetic quadrupole moment have attracted interests in atomic systems \cite{magq5,magq6}, molecules \cite{magq7,magq9}, chiral anomaly \cite{magq}, with $P$- and $T$-odds effects in atoms \cite{magq3,magq8,magq10}, Coulomb-type interactions \cite{fb}, Landau-type quantization \cite{fb2} and analogue of the quantum Hall effect \cite{fb3}. Other interesting studies have been made in Refs. \cite{magq1,magq2,magq4}. In this work, we consider the single particle approximation used in Refs. \cite{prc,pra,fb,chen}, and then we deal with a system that consists in a moving atom with a magnetic quadrupole moment that interacts with external fields. Then, we introduce the Landau quantization associated with a moving atom that possesses a magnetic quadrupole moment, and thus we analyse rotating effects on the Landau-type quantization.

The structure of this paper is: in section II, we make a brief introduction of the quantum dynamics for a moving atom with magnetic quadrupole moment that interacts with external fields in the single particle approximation proposed in Refs. \cite{prc,chen}, and then, we analyse the Landau quantization associated with an atom with a magnetic quadrupole moment in a rotating frame; in section III, we present our conclusions.

\section{Rotating effects the Landau-type quantization}

Since atoms and molecules that possess quadrupole moment have a great interest in chemical physics \cite{quad-1,quad-2,quad3,quad5,quad7,quad8,quad9,quad10,quad11,quad12,quad13,quad15,quad17,quad6,prlquad,magq5,magq6,magq7,magq9} and quantum physics \cite{b7,nonc,magq,magq3,magq8,magq10}, in this section, we investigate rotating effects on an atom/molecule with a magnetic quadrupole moment that interacts with an electric field when this interaction gives rise to an analogue of the Landau quantization \cite{landau}. By following Refs. \cite{pra,prc}, in the rest frame of the particle, the potential energy of atoms/molecules with a magnetic quadrupole moment is given by $U_{m}=-\sum_{i,\,j}M_{ij}\,\partial_{i}B_{j}$, where $\vec{B}$ is the magnetic field and $M_{ij}$ is the magnetic quadrupole tensor. On the other hand, it has been shown in Refs. \cite{fb,chen} that when an atom with a magnetic quadrupole moment moves with a velocity $v\ll c$ ($c$ is the velocity of light), then, the general expression for the Hamiltonian operator is given by (in SI units)
\begin{eqnarray}
\hat{H}_{0}=\frac{\hat{\pi}^{2}}{2m}-\vec{M}\cdot\vec{B}+V,
\label{1.1}
\end{eqnarray} 
where $m$ is the mass of the particle, $V$ is a scalar potential and the operator $\hat{\pi}$ is defined as 
\begin{eqnarray}
\hat{\pi}=\hat{p}-\frac{1}{c^{2}}\,\vec{M}\times\vec{E}.
\label{1.2}
\end{eqnarray}
In Eqs. (\ref{1.1}) and (\ref{1.2}) is defined a vector $\vec{M}$ whose components are determined by $M_{i}=\sum_{j}M_{ij}\,\partial_{j}$, where $M_{ij}$ is the magnetic quadrupole moment tensor, which is a symmetric and a traceless tensor \cite{pra,prc}. Besides, the fields $\vec{E}$ and $\vec{B}$ given in Eqs. (\ref{1.1}) and (\ref{1.2}) are the electric and magnetic fields in the laboratory frame, respectively \cite{griff}.

An analogue of the Landau quantization for a moving atom that possesses a magnetic quadrupole moment was proposed in Refs. \cite{fb2,fb3} based on the properties of the magnetic quadrupole tensor, where the field configuration in the laboratory frame that interacts with the magnetic quadrupole moment of the atom must produce a uniform effective magnetic field perpendicular to the plane of motion of the particle given by $\vec{B}_{\mathrm{eff}}=\vec{\nabla}\times\left[\vec{M}\times\vec{E}\right]$, with the vector $\vec{E}$ as being the the electric field in the laboratory frame that satisfies the electrostatic conditions. From this perspective, it has been shown in Ref. \cite{fb2} that an analogue of the Landau quantization can be obtained by considering the magnetic quadrupole moment tensor to be defined by the components: 
\begin{eqnarray}
M_{\rho z}=M_{z\rho}=M,
\label{1.3}
\end{eqnarray}
where $M$ is a constant $\left(M>0\right)$ and with all other components of $M_{ij}$ as being zero; thus, this magnetic quadrupole moment interacts with an electric field given by
\begin{eqnarray}
\vec{E}=\frac{\lambda\,\rho^{2}}{2}\,\hat{\rho},
\label{1.4}
\end{eqnarray}
where $\lambda$ is a constant associated with a non-uniform distribution of electric charges inside a non-conductor cylinder. In this particular case, we have that the magnetic quadrupole moment defined in Eq. (\ref{1.3}) is a symmetric and traceless matrix. We also have an effective vector potential given by $\vec{A}_{\mathrm{eff}}=\vec{M}\times\vec{E}=\lambda\,M\,\rho\,\hat{\varphi}$ (where $\hat{\varphi}$ is a unit vector in the azimuthal direction) and, consequently, there exists an effective magnetic field which is uniform in the $z$-direction $\vec{B}_{\mathrm{eff}}=\lambda\,M\,\hat{z}$ (where $\hat{z}$ is a unit vector in the $z$-direction), that is, it is perpendicular to the plane of motion of the quantum particle. Therefore, the conditions for achieving the Landau quantization associated with a moving atom that possesses a magnetic quadrupole moment are satisfied.

Henceforth, let us consider the Landau-type system above to be rotating with a constant angular velocity $\vec{\Omega}=\Omega\,\hat{z}$, then, it has been discussed in Refs. \cite{landau3,landau4,dantas,anan,r13} that the Hamiltonian operator that describes the behaviour of a quantum system in a rotating frame is given by 
\begin{eqnarray}
\hat{H}=\hat{H}_{0}-\vec{\Omega}\cdot\hat{L},
\label{1.5}
\end{eqnarray} 
where $\hat{L}$ is the angular momentum operator given by $\hat{L}=\vec{r}\times\hat{\pi}$, where $\hat{\pi}$ is given in eq. (\ref{1.2}). In this two-dimensional system, we can write $\vec{r}=\rho\,\hat{\rho}$, where $\rho=\sqrt{x^{2}+y^{2}}$ is the radial coordinate and $\hat{\rho}$ is a unit vector in the radial direction. In recent decades, geometric quantum phases have been investigated in nonrelativistic quantum systems \cite{anan,r14,r15,r13} based on the approach of Eq. (\ref{1.5}). For a moving atom with a magnetic quadrupole moment, hence, the Schr\"odinger equation becomes (we shall work with the units $\hbar=c=1$ from now on):
\begin{eqnarray}
i\frac{\partial\psi}{\partial t}&=&-\frac{1}{2m}\left[\frac{\partial^{2}}{\partial\rho^{2}}+\frac{1}{\rho}\,\frac{\partial}{\partial\rho}+\frac{1}{\rho^{2}}\,\frac{\partial^{2}}{\partial\varphi^{2}}+\frac{\partial^{2}}{\partial z}\right]\psi-i\frac{M\,\lambda}{m}\,\frac{\partial\psi}{\partial\varphi}+\frac{M^{2}\,\lambda^{2}}{2m}\,\rho^{2}\,\psi\nonumber\\
[-2mm]\label{1.6}\\[-2mm]
&+&i\Omega\,\frac{\partial\psi}{\partial\varphi}+M\,\lambda\,\Omega\,\rho^{2}\psi.\nonumber
\end{eqnarray}

Note that the Hamiltonian operator of the right-hand-side of Eq. (\ref{1.6}) commutes with the operators $\hat{L}_{z}=-i\,\frac{\partial}{\partial\varphi}$ and $\hat{p}_{z}=-i\,\frac{\partial}{\partial z}$, then, a particular solution can be written in terms of the eigenvalues of  $\hat{L}_{z}$ and  $\hat{p}_{z}$ as $\psi\left(t,\,\rho,\,\varphi,\,z\right)=e^{-i\mathcal{E}t}\,e^{i\,l\,\varphi}\,e^{ikz}\,R\left(\rho\right)$, where $l=0,\pm1,\pm2,\ldots$, $k$ is a constant and $R\left(\rho\right)$ is a function of the radial coordinate. From now on, we assume that $k=0$ in order to reduce the system to a planar system. By substituting this particular solution into Eq. (\ref{1.6}), we have
\begin{eqnarray}
\frac{d^{2}R}{d\rho^{2}}+\frac{1}{\rho}\frac{dR}{d\rho}-\frac{l^{2}}{\rho^{2}}-\frac{m^{2}\delta^{2}}{4}\,\rho^{2}\,R+2m\left[\mathcal{E}-\Omega\,l-\frac{l\,\omega}{2}\right]R=0,
\label{1.7}
\end{eqnarray}
where we have defined the parameters in Eq. (\ref{1.7}):
\begin{eqnarray}
\delta^{2}=\omega^{2}+4\Omega\,\omega;\,\,\,\,\,\,\omega=\frac{2\,M\,\lambda}{m}.
\label{1.8}
\end{eqnarray}
Note that the parameter $\omega$ was defined in Ref. \cite{fb2} as being the cyclotron frequency of the Landau quantization associated with an atom with a magnetic quadrupole moment. Next, by performing a change of variables given by $r=\frac{m\delta}{2}\,\rho^{2}$, hence, the Schr\"odinger equation (\ref{1.7}) becomes
\begin{eqnarray}
r\,\frac{d^{2}R}{dr^{2}}+\frac{dR}{dr}-\frac{l^{2}}{4r}\,R-\frac{r}{4}\,R+\beta\,R=0,
\label{1.9}
\end{eqnarray}
where $\beta=\frac{1}{\delta}\left[\mathcal{E}+\Omega\,l-\frac{1}{2}\,\omega\,l\right]$. By analysing the asymptotic behaviour of the possible solutions to Eq. (\ref{1.9}), which is determined for $r\rightarrow0$ and $r\rightarrow\infty$, then, we can write the function $R\left(r\right)$ in terms of an unknown function $M\left(r\right)$ as follows \cite{landau,fb2}:
\begin{eqnarray}
R\left(r\right)=e^{-r^{2}/2}\,r^{\left|l\right|/2}\,M\left(r\right),
\label{1.10}
\end{eqnarray}
and thus, by substituting Eq. (\ref{1.10}) into Eq. (\ref{1.9}), we obtain
\begin{eqnarray}
r\frac{d^{2}M}{dr^{2}}+\left[\left|l\right|+1-r\right]\frac{dM}{dr}+\left[\beta-\frac{\left|l\right|}{2}-\frac{1}{2}\right]M=0,
 \label{1.11}
\end{eqnarray}
which is called as the confluent hypergeometric equation \cite{landau,abra,arf} and $M\left(r\right)=M\left(\frac{\left|l\right|}{2}+\frac{1}{2}-\beta,\,\left|l\right|+1,\,r\right)$ is the confluent hypergeometric function. Since our focus is on the bound state solutions, then, let us impose that the confluent hypergeometric series becomes a polynomial of degree $n$, where $n=0,1,2,\ldots$. This is possible when $\frac{\left|l\right|}{2}+\frac{1}{2}-\beta=-n$, therefore we obtain
\begin{eqnarray}
\mathcal{E}_{n,\,l}=\sqrt{\omega^{2}+4\Omega\,\omega}\,\left[n+\frac{\left|l\right|}{2}+\frac{1}{2}\right]-\frac{1}{2}\,\omega\,l-\Omega\,l,
\label{1.12}
\end{eqnarray}
where $n=0,1,2,\ldots$ is the quantum number associated with the radial modes and $l=0,\pm1,\pm2,\ldots$ is the angular momentum quantum number.

Hence, the energy levels (\ref{1.12}) stems from the rotating effects on the Landau quantization associated with an atom with an magnetic quadrupole moment. Note that the cyclotron frequency of the analogue of the Landau levels is modified by the influence of the rotating effects and becomes $\delta=\sqrt{\omega^{2}+4\Omega\,\omega}$. Note that the degeneracy of the Landau levels obtained is broken due to the rotating effects. On the other hand, by taking the limit $\Omega\rightarrow0$, we recover the cyclotron frequency and the analogue of the Landau levels given in Ref. \cite{fb2}. Finally, observe that the last term of Eq. (\ref{1.12}) corresponds to the coupling between the angular momentum quantum number $l$ and the angular velocity $\Omega$ which is known in the literature as the Page-Werner {\it et al} term \cite{r1,r2,r3}.

\section{conclusions}

We have investigated rotating effects on the Landau quantization associated with an atom with a magnetic quadrupole moment. We have seen that the effects of rotation modify the cyclotron frequency and break the degeneracy of the analogue of the Landau levels. Moreover, a new contribution to the analogue of the Landau levels arises from the coupling between the angular velocity of the rotating frame and the angular momentum, which is called as the Page-Werner {\it et al} term \cite{r1,r2,r3}. Furthermore, we have seen that by taking the limit $\Omega\rightarrow0$ we recover the Landau quantization associated with an atom with a magnetic quadrupole moment \cite{fb2}.

Observe that the present study has an interest in condensed matter systems that possess linear topological defects. In Refs. \cite{kleinert,kat} is shown a geometric approach to describe topological defects in solids, such as disclinations and dislocations, based on the Riemann geometry that has been explored in the current literature. For instance, it has been observed that the presence of linear topological defects in mesoscopic systems can break the degeneracy of the Landau levels \cite{fur}. Hence, the effects of linear topological defects on the Landau quantization associated with an atom that possesses a magnetic quadrupole moment in a rotating frame is an interesting point of discussion. Another point of view that deals with rotating effects has been proposed in Refs. \cite{cond1,hall}, where the quantum Hall effect for electrically charged particles is analysed. Since the present study works with a neutral particle (atom/molecule) with a magnetic quadrupole moment, then, it differs from that of Refs. \cite{cond1,hall}, and thus, at first sight, we cannot achieve a quantum Hall system. However, other proposals for studying the quantum Hall effect for neutral particles have been proposed in recent decades \cite{er,lin5}, which can be a way of investigating the quantum Hall effect for an atom/molecule with a magnetic quadrupole moment under the effects of rotation. We hope to bring these discussions in the near future.

\acknowledgments

The authors would like to thank the Brazilian agencies CNPq and CAPES for financial support.

\end{document}